\documentstyle[prl,aps,multicol,epsf]{revtex}
\input psfig

\def\be{\begin{equation}}
\def\ee{\end{equation}}

\def\sqr#1#2{{\vcenter{\vbox{\hrule height .#2pt
      \hbox{\vrule width .#2pt height#1pt \kern#1pt
      \vrule width.#2pt}
      \hrule height.#2pt}}}}
\def\square{\mathchoice\sqr34\sqr34\sqr{2.1}3\sqr{1.5}3}

\begin{document}
\bibliographystyle{simpl1}

\title{A Consistent Treatment of Strong Coupling in Disordered
Superconducting Films}

\author{Margit Steiner and Robert A Smith}

\address{School of Physics and Astronomy, University of
Birmingham, Edgbaston, Birmingham B15 2TT, ENGLAND}

\maketitle
\bigskip

\begin{abstract}
We develop a consistent treatment of disorder effects in strong coupling
superconductors. We use two different approaches, starting either from above
or below the transition temperature, and show their equivalence. The normal
state approach is a natural extension of a recently developed non-perturbative 
resummation technique. For the superconducting state we extend the standard
Eliashberg theory to include disorder effects. We obtain a compact set of
equations which are physically transparent and easy to use. They correctly
reproduce the usual strong coupling results and disordered perturbation
theory results in the appropriate limits. We numerically solve these equations
to obtain transition temperature as a function of disorder for different
phonon couplings.  
\end{abstract}
\bigskip

\begin{multicols}{2}

The phenomenon of the suppression of the transition temperature in
thin superconducting films by disorder has been extensively investigated
in recent years \cite{Fink94}. The physical mechanism is easy to understand: disorder
causes electrons to move diffusively, leading to a less efficient screening
of the Coulomb repulsion between them. This manifests itself as a 
low-momentum singularity in the dynamically screened Coulomb interaction.
Naively one would expect this to lead to a very strong suppression of
superconductivity due to the enhancement of the effective Coulomb repulsion 
\cite{Fuk84}.
However this turns out not to be the case because of a cancellation
between the various singular terms which arise, for example,
in perturbation theory \cite{Fink87}.
This apparently fortuitous cancellation is in fact due to gauge invariance
\cite{SA00}, and
the screened Coulomb potential can therefore be treated as if it were
featureless. We note that all these subtleties are present even in the
simplest possible calculation, namely first order perturbation theory
assuming a featureless BCS attraction. However this model is too restrictive
for the following reasons: (i) perturbation theory is not valid in the
limit of strong disorder often seen experimentally; (ii) many of the
experimentally relevant materials such as Pb
and Pb-Bi are strong coupling superconductors
for which the details of the phonon spectrum are important 
\cite{XHD,VDG,CV,Scal69,AM82}. It is therefore
desirable to extend the simple theory to accomodate these features in a
consistent manner.

A method of going beyond perturbation theory has recently been developed by
Oreg and Finkel'stein\cite{OF}. These authors use a ladder summation technique which
treats the BCS attraction and impurity-dressed Coulomb repulsion on an equal
footing. This is possible because both interactions can be treated as featureless,
as explained above. The fact that this technique reproduces the renormalization
group equation known from the interacting non-linear $\sigma$-model
\cite{Fink83} proves its
essential correctness. Its great advantage is its transparency and the fact
that it is easy to use. The aim of this paper is to extend this technique to
include strong coupling effects. Initially one might think that this would
further complicate the problem. Counterintuitively the problem actually becomes
simpler because one realises that the original technique is best understood
in terms of strong coupling theory. Moreover this realisation allows us to
naturally extend the non-perturbative treatment of disorder to below the 
transition temperature. 

In this paper we present a consistent treatment of the effects of 
disorder and strong coupling on the transition temperature in superconductors. 
We use two different approaches which we show to be equivalent: 
(a) Starting from the normal state we derive 
the pair propagator and identify the transition temperature as the 
temperature at which it becomes singular; (b) following the Eliashberg theory
of strong-coupling superconductors, we start from below 
the transition temperature and we derive the superconducting self-energy and
identify the transition temperature as the point at which the 
order parameter goes to zero. The former approach has the advantage that 
it does not make any assumptions about the structure of the superconducting 
state and so is the natural way to initially address the problem. 
The latter approach has the advantage that it allows us to gain information
about the superconducting state at all temperatures below the transition 
temperature at no extra cost. The equivalence of the results obtained by the
two approaches justifies the assumptions we make about the superconducting
state.

We begin by calculating the zero-momentum pair propagator, 
$\Gamma(\omega,\omega')$, shown in Fig. 1. This yields the 
equation
%%%%%%%%%%%%%%%%%%%%%%%%%%%%%%%%%%%%%%%%%%%%%%%%%%%%%%%%%%%%%%%%%%%%%
\begin{equation}
\label{Gammadef}
\hspace{-0.15cm}\Gamma(\omega,\omega')\hspace{-0.05cm}= 
\hspace{-0.05cm}\Gamma_0(\omega,\omega')\hspace{-0.05cm}+\hspace{-0.05cm}
T \sum_{\omega''} \Gamma_0(\omega,\omega'') C(\omega'')
\Gamma(\omega'',\omega')
\end{equation}
%%%%%%%%%%%%%%%%%%%%%%%%%%%%%%%%%%%%%%%%%%%%%%%%%%%%%%%%%%%%%%%%%%%%
where the $\omega$ are Fermi Matsubara frequencies,
$\Gamma_0$ is the elementary interaction vertex and $C$ is the 
Cooperon impurity ladder.  
All momentum dependences have been removed by averaging over the 
Fermi surface. 
We identify the transition temperature as the temperature at which $\Gamma$
diverges. This will occur when there is an eigenvector $\phi(\omega)$
of the homogeneous part of Eq. (\ref{Gammadef}), which leads to the equation
%%%%%%%%%%%%%%%%%%%%%%%%%%%%%%%%%%%%%%%%%%%%%%%%%%%%%%%%%%%%%%%%%
\begin{eqnarray}
\label{Gammaeig}
\lefteqn{
T \sum_{\omega'} \Gamma(\omega,\omega') \phi(\omega')=}\nonumber\\
&&\hspace{1cm}
T \sum_{\omega'} T \sum_{\omega''} \Gamma_0(\omega,\omega'') 
C(\omega'') \Gamma(\omega'',\omega') \phi(\omega').
\end{eqnarray}
%%%%%%%%%%%%%%%%%%%%%%%%%%%%%%%%%%%%%%%%%%%%%%%%%%%%%%%%
If we define $\Delta(\omega) = T \sum_{\omega'}
\Gamma(\omega,\omega') \phi(\omega')$ the above equation has the 
familiar Eliashberg form for the order parameter,
%%%%%%%%%%%%%%%%%%%%%%%%%%%%%%%%%%%%%%%%%%%%%%%%%%%%%%%%
\begin{equation}
\label{Deltadef}
\Delta(\omega) = T \sum_{\omega''} \Gamma_0(\omega,\omega'') 
C(\omega'') \Delta(\omega'').
\end{equation}
%%%%%%%%%%%%%%%%%%%%%%%%%%%%%%%%%%%%%%%%%%%%%%%%%%%%%%%%
 
The elementary interaction vertex $\Gamma_0$ contains the electron-phonon
contribution, $\Gamma_{0,ph}$, and the Coulomb contribution. The latter naturally
splits into two parts: the bare Coulomb interaction, $\Gamma_{0,C}$, for high
momentum and frequency transfer, and the impurity-dressed Coulomb interaction,
$\Gamma_{0,dis}$ otherwise. The second contribution
is important for low momentum transfers, $|q|<1/\ell$, and low frequency transfers,
$|\omega|<1/\tau$, where $\ell$ is the elastic mean free path and $\tau$ is the
elastic scattering time.  This is because the disorder-screened Coulomb 
interaction keeps its long-range
momentum singularity and its effective magnitude is therefore greatly enhanced.
Gauge invariance leads to a cancellation of singular terms \cite{Fink87,SRW,SA00} 
which allows us to
treat the disorder-screened interaction as featureless with magnitude
$V_{dis}=1/2N(0)$, where $N(0)$ is the electronic density of states at the Fermi
surface.

The electron-phonon contribution, $\Gamma_{0,ph}$, is given by
%%%%%%%%%%%%%%%%%%%%%%%%%%%%%%%%%%%%%%%%%%%%%%%%%%%%%%%%%%%%%%%%%
\begin{eqnarray}
\label{Gammaph}
\Gamma_{0,ph}(\omega,\omega') & = & \sum_\lambda \mbox{\large $\langle$}
|g_{{\bf k},{\bf k'},\lambda}|^2 D_\lambda({\bf k -k'},\omega - \omega')
\mbox{\large $\rangle$}_{FS} \nonumber \\
&&\hspace{-1.2cm} = \frac{1}{N(0)} \int_0^\infty dz \,
\frac{2 z \alpha^2(z) F(z)}{(\omega - \omega')^2 + z^2}
\equiv {\lambda(\omega-\omega')\over N(0)},
\end{eqnarray}
%%%%%%%%%%%%%%%%%%%%%%%%%%%%%%%%%%%%%%%%%%%%%%%%%%%%%%%%
where we have averaged over the Fermi surface in the standard manner.
In the above equation $g_{{\bf k},{\bf k'},\lambda}$ is the 
electron-phonon coupling for phonons of polarisation $\lambda$, 
$D_\lambda({\bf k -k'},\omega - \omega')$ is the phonon Green function
and 
$\alpha^2(z) F(z) = N(0) \sum_\lambda \mbox{\large $\langle$}  
|g_{{\bf k},{\bf k'},\lambda}|^2 \delta(z-\omega_{{\bf k -k'},\lambda})
\mbox{\large $\rangle$}_{FS}$. The bare Coulomb contribution, $\Gamma_{0,C}$,
 is just the Coulomb potential averaged over the Fermi surface,
%%%%%%%%%%%%%%%%%%%%%%%%%%%%%%%%%%%%%%%%%%%%%%%%%%%%%%%%%%%%%%%%%
\begin{equation}
\label{Gammacoul}
\Gamma_{0,C}=V_C=\mbox{\large $\langle$}V({\bf k}-{\bf k'})
\mbox{\large $\rangle$}_{FS}. 
\end{equation}
%%%%%%%%%%%%%%%%%%%%%%%%%%%%%%%%%%%%%%%%%%%%%%%%%%%%%%%%%%%%%%%%%
The disordered Coulomb contribution is given by
%%%%%%%%%%%%%%%%%%%%%%%%%%%%%%%%%%%%%%%%%%%%%%%%%%%%%%%%%%%%%%%%%
\begin{equation}
\label{Gammadis}
\Gamma_{0,dis}
=\pi t\log{\left({1\over (|\omega|+|\omega'|)\tau}\right)}
\end{equation}
%%%%%%%%%%%%%%%%%%%%%%%%%%%%%%%%%%%%%%%%%%%%%%%%%%%%%%%%%%%%%%%%%
where we define the dimensionless measure of disorder,
$t=V_{dis}/4\pi^2D=R_{\square}/R_0$, and $R_0=2\pi h/e^2=162k\Omega$
is a quantum unit of resistance.

First we include only the bare Cooperon ladder, $C_0=\pi N(0)/|\omega|$,
to yield the self-consistency equation
%%%%%%%%%%%%%%%%%%%%%%%%%%%%%%%%%%%%%%%%%%%%%%%%%%%%%%%%%%%%%%%%%%%%%%%
\begin{figure}[t]
\centerline{\psfig{figure=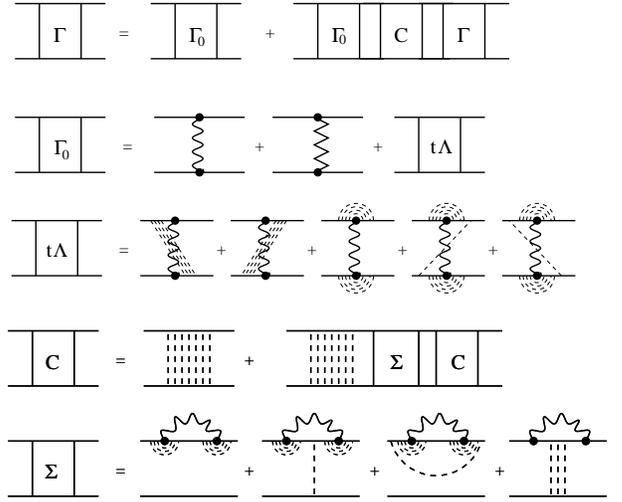,width=8cm}}
\medskip
\caption{Diagrammatic equation for the pair propagator $\Gamma$ in terms
of the elementary interaction vertex $\Gamma_0$ and the screened
Cooperon $C$. $\Gamma_0$ consists of electron-phonon interaction
(zig-zag line), the bare Coulomb interaction
(wiggly line)  and
the disorder-dressed Coulomb interaction block $t \Lambda$. $\Sigma$
is the self-energy correction to the Cooperon due to the
impurity-screened Coulomb interaction.}
\end{figure}
%%%%%%%%%%%%%%%%%%%%%%%%%%%%%%%%%%%%%%%%%%%%%%%%%%%%%%%%%%%%%%%%%%%%%%%
%%%%%%%%%%%%%%%%%%%%%%%%%%%%%%%%%%%%%%%%%%%%%%%%%%%%%%%%%%%%%%%%%
\begin{eqnarray}
\label{Delta}
\lefteqn{\Delta(\omega)=\pi T\sum_{\omega''} {\Delta(\omega'')\over |\omega''|}
\bigg[\lambda(\omega-\omega'')}\nonumber\\
&&\hspace{2.5cm}
-\mu-t\ln{\left({1\over (|\omega|+|\omega''|)\tau}\right)}\bigg],
\end{eqnarray}
%%%%%%%%%%%%%%%%%%%%%%%%%%%%%%%%%%%%%%%%%%%%%%%%%%%%%%%%%%%%%%%%%
\noindent
where $\mu=N(0)V_C$.
The sums over the Matsubara frequencies in the above equation are
cut off at the Fermi energy. However the phonon interaction is
effective only for energies of the order the Debye frequency. 
We therefore follow the standard procedure \cite{AM82} of introducing a frequency
$\omega_c$ above which only the bare Coulomb term contributes. The
equation for $\Delta(\omega)$ with $\omega>\omega_c$ then decouples
and the only effect of these frequencies is to renormalize $\mu$ to
%%%%%%%%%%%%%%%%%%%%%%%%%%%%%%%%%%%%%%%%%%%%%%%%%%%%%%%%%%%%%%%%%
\begin{equation}
\label{UC}
\mu^*={\mu\over 1+\mu\displaystyle\ln{(\epsilon_F/\omega_c)}}.
\end{equation}
%%%%%%%%%%%%%%%%%%%%%%%%%%%%%%%%%%%%%%%%%%%%%%%%%%%%%%%%%%%%%%%%%
We must therefore replace $\mu$ by $\mu^*$ in Eq. (\ref{Delta}), and cut off
the sums at frequency $\omega_c$.

Next we consider self-energy corrections to the impurity-dressed electron
Green function. Having averaged out the momentum dependence, the self-energy
has the form $\Sigma(\omega)=i\omega[1-Z(\omega)]$. The dressing of the electron
Green function leads to the Cooperon being modified to 
$C(\omega)=\pi N(0)/(Z(\omega)|\omega|)$, as shown in Fig. (1).
At this point we note the difference between the effect of including the
impurity self-energy in the original electron Green function,
and the effects of including the phonon and Coulomb terms. The impurity 
correction, $\Sigma_{imp}=(i/2\tau)sgn(\omega)$, is exactly cancelled 
in the process of
constructing the impurity ladder -- this is essentially the content of Anderson's
theorem \cite{And59} which says that non-magnetic impurities have no effect on 
superconductivity at the mean field level. The contributions to the self-energy
are   
%%%%%%%%%%%%%%%%%%%%%%%%%%%%%%%%%%%%%%%%%%%%%%%%%%%%%%%%%%%%%%%%%
\begin{eqnarray}
\label{Selfen}
\lefteqn{i\omega(1-Z(\omega))=
-i\pi T\sum_{\omega'}\lambda(\omega-\omega')sgn(\omega')}\nonumber\\
&&\hspace{3cm}+it\omega\pi T\sum_{\omega'} {1\over |\omega|+|\omega'|}.
\end{eqnarray}
%%%%%%%%%%%%%%%%%%%%%%%%%%%%%%%%%%%%%%%%%%%%%%%%%%%%%%%%%%%%%%%%%
The presence of $Z(\omega)$ in the denominator of $C(\omega)$ means that
we should redefine 
$\Delta(\omega)=T\sum_{\omega'}\Gamma(\omega,\omega')\phi(\omega')/Z(\omega)$
in Eq. (\ref{Gammaeig})
so that $\Delta(\omega)$ is given by
%%%%%%%%%%%%%%%%%%%%%%%%%%%%%%%%%%%%%%%%%%%%%%%%%%%%%%%%%%%%%%%%%
\begin{eqnarray}
\label{ZDelta}
\lefteqn{Z(\omega)\Delta(\omega)=
\pi T\sum_{\omega''<\omega_c} {\Delta(\omega'')\over |\omega''|}
\bigg[\lambda(\omega-\omega'')}\nonumber\\
&&\hspace{3cm}
-\mu^*-t\ln{\left({1\over (|\omega|+|\omega''|)\tau}\right)}\bigg]
\end{eqnarray}
%%%%%%%%%%%%%%%%%%%%%%%%%%%%%%%%%%%%%%%%%%%%%%%%%%%%%%%%%%%%%%%%%
Equations (\ref{Selfen}) and (\ref{ZDelta}) constitute the main result of
this paper. They are the Eliashberg equations for strong coupling superconductors
including the effects of Coulomb interaction and disorder.

The above equations have been derived using the pair propagator approach,
whereas they are normally derived in the framework of the Nambu-Gor'kov
self-energy formalism. We will now extend the latter formalism to include the
effects of disorder. The various contributions to the superconducting self-energy,
%%%%%%%%%%%%%%%%%%%%%%%%%%%%%%%%%%%%%%%%%%%%%%%%%%%%%%%%%%%%%%%%%%%%%%%%%%%%%%%%%
\begin{equation}
\Sigma(\omega)=i\omega[1-Z(\omega)]\tau_0
+Z(\omega)\Delta(\omega)\tau_1,
\end{equation}
%%%%%%%%%%%%%%%%%%%%%%%%%%%%%%%%%%%%%%%%%%%%%%%%%%%%%%%%%%%%%%%%%%%%%%%%%%%%%%%%%
%%%%%%%%%%%%%%%%%%%%%%%%%%%%%%%%%%%%%%%%%%%%%%%%%%%%%%%%%%%%%%%%%%%%%%%%%%%%%%%%%
\begin{figure}[t]
\centerline{\psfig{figure=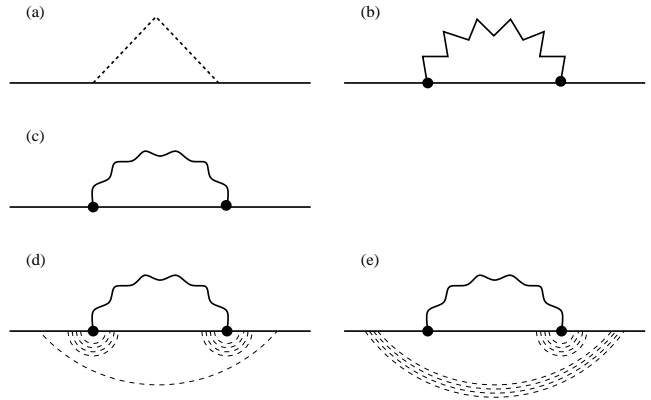,width=8.5cm}}
\medskip
\caption{Contributions to the superconducting self-energy due to:
(a) impurity scattering, (b) electron-phonon interaction, (c)
bare Coulomb repulsion, (d) and (e) disorder-dressed Coulomb interaction.}
\end{figure}
%%%%%%%%%%%%%%%%%%%%%%%%%%%%%%%%%%%%%%%%%%%%%%%%%%%%%%%%%%%%%%%%%%%%%%%%%%%%%%%%%
\noindent
are shown in Fig. (2). The phonon and bare Coulomb contributions are
%%%%%%%%%%%%%%%%%%%%%%%%%%%%%%%%%%%%%%%%%%%%%%%%%%%%%%%%%%%%%%%%%%%%%%%%%%%%%%%%%
\begin{eqnarray}
\Sigma_{ph}(\omega)&=&-\pi T\sum_{\omega'}\lambda(\omega-\omega')
{i\omega'\tau_0-\Delta(\omega')\tau_1\over W'}\nonumber\\
\Sigma_{C}(\omega)&=&U_C\pi T\sum_{\omega'}{\Delta(\omega')\over W'},
\end{eqnarray}
%%%%%%%%%%%%%%%%%%%%%%%%%%%%%%%%%%%%%%%%%%%%%%%%%%%%%%%%%%%%%%%%%%%%%%%%%%%%%%%%%
where $W=\sqrt{\omega^2+\Delta^2(\omega)}$ and 
$W'=\sqrt{\omega'^2+\Delta^2(\omega')}$.
They can be found in Ref. \cite{AM82}.
The disorder-dressed Coulomb diagrams of Fig. (2d) and (2e) give
%%%%%%%%%%%%%%%%%%%%%%%%%%%%%%%%%%%%%%%%%%%%%%%%%%%%%%%%%%%%%%%%%%%%%%%%%%%%%%%%%
\end{multicols}
\noindent
\rule{0.5\textwidth}{0.4pt}\rule{0.4pt}{\baselineskip}
\begin{eqnarray}
\Sigma_{dis,1}(\omega)&=&-t\pi T\sum_{\omega'}
{i\omega\tau_0-\Delta(\omega)\tau_1\over (W+W')}
\left[1-{\omega\omega'+\Delta(\omega)\Delta(\omega')\over WW'}\right]
\nonumber\\
\Sigma_{dis,2}(\omega)&=&t\pi T\sum_{\omega'}
\ln{\left[{1\over (W+W')\tau}\right]}
\left({\omega'\Delta(\omega)-\omega\Delta(\omega')\over WW'}\right)
\left({i\Delta(\omega)\tau_0+\omega\tau_1\over W}\right).
\end{eqnarray}
\hspace*{\fill}\rule[0.4pt]{0.4pt}{\baselineskip}%
\rule[\baselineskip]{0.5\textwidth}{0.4pt}
\begin{multicols}{2}
%%%%%%%%%%%%%%%%%%%%%%%%%%%%%%%%%%%%%%%%%%%%%%%%%%%%%%%%%%%%%%%%%%%%%%%%%%%%%%%%%
In the limit where $T\rightarrow T_c$, the order parameter 
$\Delta(\omega)\rightarrow 0$, and we obtain the previously derived Eliashberg
equations (\ref{Selfen}) and (\ref{ZDelta}).

Having derived the Eliashberg equations in Matsubara frequencies, let us now
proceed to their numerical solution. Following Refs. \cite{AM82,AD75}, we multiply
Eq. (\ref{Selfen}) by $\Delta(\omega)$ and equate it to Eq. (\ref{ZDelta}).
Noting that $\omega=(2n+1)\pi T$, $\omega'=(2n'+1)\pi T$, and defining
$\tilde{\Delta}_n=\Delta(\omega)/(2n+1)^{1/2}$, we obtain the matrix
equation
%%%%%%%%%%%%%%%%%%%%%%%%%%%%%%%%%%%%%%%%%%%%%%%%%%%%%%%%%%%%%%%%%%%%%%%%%%%%%%%%%
\begin{equation}
\sum_{n'=0}^{n_c} {S}_{nn'}\tilde{\Delta}_{n'}=0,
\end{equation}
%%%%%%%%%%%%%%%%%%%%%%%%%%%%%%%%%%%%%%%%%%%%%%%%%%%%%%%%%%%%%%%%%%%%%%%%%%%%%%%%%
where
%%%%%%%%%%%%%%%%%%%%%%%%%%%%%%%%%%%%%%%%%%%%%%%%%%%%%%%%%%%%%%%%%%%%%%%%%%%%%%%%%
\end{multicols}
\noindent
\rule{0.5\textwidth}{0.4pt}\rule{0.4pt}{\baselineskip}
\begin{eqnarray}
{S}_{nn'}&=&\left\{\lambda(n-n')+\lambda(n+n'+1)-2\mu^*-
2t\ln{\left[{M\over n+n'+1}\right]}\right\}
{1\over[(2n+1)(2n'+1)]^{1/2}}\nonumber\\
&-&\delta_{nn'}\left\{1+\left(\lambda(0)+2\sum_{n''=1}^n \lambda(n'')\right)
{1\over[(2n+1)(2n'+1)]^{1/2}}+t\sum_{n''=0}^{n_c}{1\over n+n''+1}\right\}.
\end{eqnarray}
\begin{multicols}{2}
%%%%%%%%%%%%%%%%%%%%%%%%%%%%%%%%%%%%%%%%%%%%%%%%%%%%%%%%%%%%%%%%%%%%%%%%%%%%%%%%%
\begin{figure}[tb]
\centerline{\psfig{figure=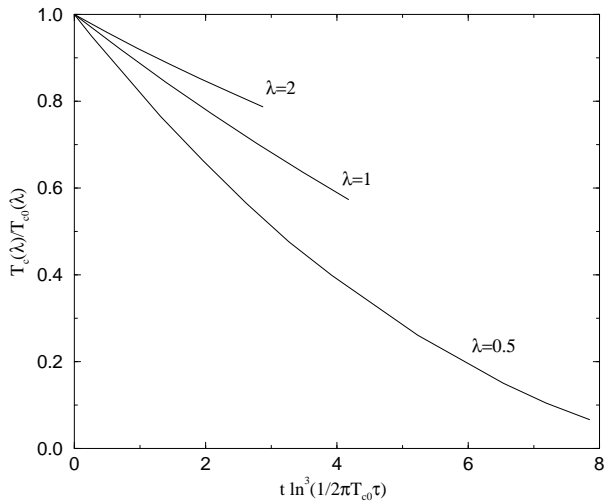,width=8cm}}
\medskip
\caption{Relative suppression of the transition temperature $T_c$ vs
$t \ln^3(1/2 \pi T_{c,0} \tau)$ for different values
of the electron-phonon coupling $\lambda$. All calculations are performed
using an Einstein spectrum.}
\end{figure}
\noindent
and $M=1/2\pi T_c\tau$.
The numerical strategy is to find the temperature at which the largest eigenvalue 
of the matrix $S$ becomes positive. We have to specify the phonon elements,
$\lambda(n)$, the Coulomb pseudopotential, $\mu^*$, the effective disorder
strength, $t=R_{\square}/R_0$, and the elastic scattering rate, $1/\tau$.
As an illustration, we solve the equation for an Einstein phonon spectrum,
$\alpha^2F(z)=\lambda\omega_E\delta(z-\omega_E)/2$, for which
$\lambda(n)=\lambda\omega_E^2/[\omega_E^2+(2n\pi T)^2]$. We plot
$T_c(\lambda,t)/T_c(\lambda,0)$ vs $t\ln^3{(1/2\pi T_c(\lambda,0)\tau)}$,
the latter being the initial slope predicted from perturbation theory.
This enables us to see if the shape of the curve deviates from the weak coupling
prediction as $\lambda$ is increased. The only parameters we fix are $\mu^*=0.1$
and $1/\omega_E\tau=250$ which are typical values for lead films. The results
are shown in Fig. 3. We see that there is a significant deviation from the
weak coupling predictions which may be important in fitting experimental data.
We intend to extend our numerical analysis to lower values of $\lambda$ and
different phonon spectra.

In conclusion we have developed a consistent treatment of the effect of
disorder and strong coupling on the transition temperature of superconducting
films. Previous authors have considered similar systems within an exact eigenstate
framework \cite{Bel87}, but this approach remains somewhat opaque physically,
and its relation to weak coupling theories is unclear. Our approach has the
advantage of yielding a compact set of equations that are easy to use, and whose
weak coupling and weak disorder limits reproduce the known results. In future work
we will analytically continue these equations to real frequencies, and use them to
analyse the effect
of disorder on tunneling characteristics in strong coupling materials.

\bigskip
\centerline {\bf ACKNOWLEDGEMENTS}
\medskip

We thank V. Ambegaokar, A.M. Finkel'stein and Y. Oreg for useful discussions.
RAS acknowledges support from the Nuffield Foundation.
MS is supported by the UK EPSRC under Grant No. RRAH06594.

\end{multicols}

\end{document}